\documentstyle{article}

\title{\bf Comments on the 2nd order bootstrap relation}
\author{M.A.Braun  \\
 Department of high-energy physics, University of S. Petersburg}

\def\beq{\begin{equation}}
\def\eeq{\end{equation}}
\def\noi{\noindent}
\def\oq{\omega(q)}
\def\oa{\omega(q_{1})}
\def\ob{\omega(q_{2})}
\def\eq{\eta (q)}
\def\ea{\eta (q_{1})}
\def\eb{\eta (q_{2})}
\def\ec{\eta (q'_{1})}
\def\ed{\eta (q'_{2})}
\def\ee{\eta(k)}
\def\xq{\xi(q)}
\def\xa{\xi(q_1)}
\def\xb{\xi(q_2)}
\def\xc{\xi(q'_1)}
\def\xd{\xi(q'_2)}
\def\xe{\xi(k)}

\begin{document}
\maketitle
\medskip
\noi{\bf Abstract.}
The 2nd order bootstrap relation is discussed in view of the recent
critics by F.Fadin, R.Fiore y A.Papa. It is shown that the
strong bootstrap
condition and the anzatz to solve it used in our earlier paper
are valid at least for the quark part of the next-to-leading
contribution.

\section{ Introduction.}
The bootstrap relation plays a key role in the derivation of the
BFKL equation up to the 2nd order in $\alpha_s$. It guarantees that
production amplitudes with the gluon quantum number in their $t$
channels used for the construction of the absorptive part are indeed
given only by a single reggeized gluon exchange and do not contain
admixture from two or more reggeized gluon exchanges.
The bootstrap relation is known to be satisfied in the lowest order
in the coupling constant. Recently  the 2nd order
bootstrap relation was discussed in [1].
In a stronger form it was used in [2] to obtain the potential in the
$t$-channel with the gluon colour quantum number. This approach was
lately critisized in [3], where it was claimed that the strong bootstrap
condition used in [2] was not fulfilled and the anzatz used to solve it
was incorrect. 
In this note we demonstrate that these objections are totally unfounded.
They are
a result of  a misinterpretation of the potential used in [2], which is
a different quantity as compared to the kernel used in [1,3].
We also derive the 2nd order bootstrap relation of [1] in a simpler way
and
comment on its implication for particle-reggeon scattering amplitudes.

\section{Formalism}
To introduce the bootstrap relation in a general form it is convenient
to use an operator formalism for the (non-forward) two gluon equation.
Let the two gluons with momenta $q_{1,2}$ be described by a wave function
$\Psi(q_1,q_2)$. The total momentum $q=q_1+q_2$ will always be conserved,
so that in future $q_2=q-q_1$ and the dependence on $q$ and thus $q_2$
will be suppressed.
To facilitate comparison with [1,3] we introduce a metric
in which the scalar product of two wave functions is given
by
\beq
\langle\Psi_1|\Psi_2\rangle\equiv\int d^{D-2}q_1
\Psi^*(q_1)\Psi(q_1),
\eeq
where $D=4+2\epsilon$, $\epsilon\rightarrow 0$, is the dimension used to
regularize all the
following expressions in the infrared region. 

In this metric, up to  order $\alpha^3_s$, the absorptive part of the
scattering amplitude coming
from the two-reggeized-gluon exchange is given by
\beq
{\cal A}_j=\langle\Phi_p|G(E)|\Phi_t\rangle
\eeq
Here the  functions $\Phi_{p,t}$ are the so called impact factors
 which represent
the coupling of the external particles ($t$ from the target and $p$
from the projectile) to the two exchanged gluons.
 They are low energy
 particle-reggeon scattering amplitudes.
They have order $\alpha_s=g^2/4\pi$. 
The Green function $G(E)$
with $E=1-j$ is defined as
\beq
  G(E)=(H-E)^{-1}
\eeq
where $H$ is the two gluon (Hermithean) Hamiltonian. Its explicit
expression in the momentum representation is
\beq
\langle q_1|H|q'_1\rangle=-\delta^{D-2}(q_1-q'_1)(\omega(q_1)+
\omega(q_2))-V^{(R)}(q_1,q'_1)
\eeq
where $1+\omega(q_{1(2)})$ is the Regge trajectory of the first 
(second) gluon and $V^{(R)}$ represents the gluonic interaction
for the $t$-channel with the colour quantum number $R$.
In the lowest (1st) order in $\alpha_s$ one has [4]
\beq
\omega^{(1)}(q)=-aq^2\langle\chi|\chi\rangle
\eeq
where 
\beq
\chi(q_1)=1/\sqrt{q_1^2q_2^2},
\eeq
\beq
a=\frac{g^2N}{2(2\pi)^{D-1}}
\eeq
and $N$ is the number of colours.
 As to the gluonic interaction,
in the 1st order
\beq
V^{(R)}(q_1,q'_1)=\frac{ac_R}{\sqrt{q_1^2q_2^2{q'_1}^2{q'_2}^2}}
\left(\frac{q_1^2{q'_2}^2+q_2^2{q'_1}^2}
{k^2}-q^2\right)
\eeq
with $k=q_1-q'_1$
(the 1st denominator appears due to our metric (1)) 
The coefficient $c_R$ depends on the colour quantum number of the
$t$-channel. For the vacuum channel ($R=v$) $c_v=2$. For the channel
with
the gluon colour quantum number ($R=g$) $c_g=1$. 

In any colour channel the Green function $G(E)$ can be represented via
the solutions of the homogeneous Schroedinger equation
\beq
H\Psi_n=E_n\Psi_n.
\eeq
These solution may be taken orthonormalized
\beq
\langle\Psi_m|\Psi_n\rangle=\delta_{mn}
\eeq
(of course index $n$ may be continuous). They are assumed to form a
complete set
\beq
\sum_n|\Psi_n\rangle\langle\Psi_n|=1.
\eeq
In our metric the last equation means in the momentum space
\beq
\sum_n\Psi_n(q_1)\Psi^*_n(q'_1)=\delta^{D-2}(q_1-q'_1).
\eeq
In terms of the eigenfunctions $\Psi_n$
\beq
G(E)=\sum_n\frac{|\Psi_n\rangle\langle\Psi_n|}{E_n-E}
\eeq
and as a consequence the absorptive part ${\cal A}_j$
is expressed as
\beq
{\cal A}_j=
\sum_n\frac{\langle\Phi_p|\Psi_n\rangle\langle\Psi_n|\Phi_t
\rangle}{E_n-E}.
\eeq

Note that sometimes it is convenient to consider not the absorptive part
of the amplitude for real particles but amplitudes for particle-reggeon
scattering. One can evidently define two such amplitudes $\Psi_{p,t}$
depending on which real particle is taken, from the projectile or target.
Formally we define
\beq
\Psi_t(E)=G(E)|\Phi_t\rangle.
\eeq
It satisfies an inhomogeneous Schroedinger equation
\beq
(H-E)\Psi_t(E)=\Phi_t.
\eeq
The absorptive part can then be  expressed as
\beq
{\cal A}_j=\langle\Phi_p|\Psi_t(E)\rangle.
\eeq
In terms of the eigenfunctions $\Psi_n$ one can express
\beq
\Psi_t(E)=
\sum_n\frac{\Psi_n\langle\Psi_n|\Phi_t
\rangle}{E_n-E}.
\eeq
The other function $\Psi_p(E)$ can be introduced in a similar manner.

\section{The bootstrap conditions}
Now we concentrate on the  $t$-channel with a colour quantum number of
the gluon ($R=g$). The ideology of the BFKL equation is based on the idea
that physical amplitudes in this channel have the asymptotical behaviour
corresponding to an exchange of a single reggeized gluon. The important
requirement is thus that no contribution should come from the exchange of
two or more reggeized gluons (in the leading and next-to-leading orders).
This requirement seems to be automatically fulfilled due to the Gribov's
rule of signature conservation. Indeed the two-gluon exchange should have
an overall  positive signature and thus cannot contribute to the
amplitude with the $t$ channel corresponding to the gluon colour quantum
number, which has a negative signature. The bootstrap relation is just
a technical tool to implement this requirement.

In order that the amplitude should have the asymptotic behaviour
corresponding to a single reggeized gluon exchange without admixture
of any other terms, its absorptive part, considered as a function
of the complex angular momentum $j$, should have a simple pole at
$j=1+\omega(q)$ and no other singularities. From the representation (13~)
we conclude that

First, the spectrum of $H$ in the considered $t$ channel should contain
an eigenvalue $E_0=-\omega(q)$ with the corresponding eigenfunction
$\Psi_0$
\beq
(H+\omega(q))\Psi_0=0
\eeq

Second, in the sum over $n$ only the contribution of this eigenfunction
$\Psi_0$
should be present. All other terms should be equal to zero. Strictly
speaking, this means that both impact factors $\Phi_{p,t}$ should be
orthogonal to all other eigenfunctions $\Psi_n$, $n>0$. Due to the
completeness of the whole set it means that both $\Phi_p$ and
$\Phi_t$ should coincide with $\Psi_0$, up to a normalization factor.

This second condition  looks to be very stringent. Note that it should
be fulfilled for {\it any} projectile and/or target. So, literally
understood, it means that coupling of any particle to a pair of reggeons
is essentially the same function of the reggeonic momenta. 
As we shall see below, in fact, this condition is far weaker, due to the
fact that all our arguments are limited to the two first orders of the
perturbation theory.

There is finally a third bootstrap condition which is  a 
relation for the impact factor. With only the state $\Psi_0$ 
contributing we get from (13)
\beq
{\cal A}_j=
\frac{\langle\Phi_p|\Psi_0\rangle\langle\Psi_0|\Phi_t
\rangle}{j-1-\omega(q)}
\eeq 
Integrating this over $j$ with a proper signature factor we find the
corresponding amplitude as a function of $s$
\beq
{\cal A}(s)=-s\left(\frac{q^2}{s_0}\right)^{\omega(q)}
\frac{
\langle\Phi_p|\Psi_0\rangle\langle\Psi_0|\Phi_t
\rangle}{\sin \pi\omega(q)}\left[\left(\frac{-s}{q^2}\right)^{\omega(q)}
+\left(\frac{s}{q^2}\right)^{\omega(q)}\right]
\eeq
This should be compared with the standard form of the contribution
of a Reggeized gluon, with a scale given by $q^2$,
\beq
{\cal A}(s)=-\Gamma_p(q)\Gamma_t(q)\frac{s}{q^2}
\left[\left(\frac{-s}{q^2}\right)^{\omega(q)}
+\left(\frac{s}{q^2}\right)^{\omega(q)}\right]
\eeq
Here $\Gamma$'s represent the coupling of the projectile or target
particles to the exchanged reggeized gluon ("particle-particle-Reggeon
vertices", in the terminology of [1]). Comparing (21) and (22) we find a 
relation which should be satisfied both for the projectile and target
\beq
\Gamma(q)=
q\left(\frac{q^2}{s_0}\right)^{\omega(q)/2}
\frac{\langle\Phi|\Psi_0\rangle}{\sqrt{|\sin \pi\omega(q)|}}
\eeq
In fact the impact factor $\Phi$ itself involves the vertex $\Gamma$
(in the lowest order it is completely expressed through it). So Eq. (23)
is a non-trivial condition to be satisfied for the coupling of any 
particle to the reggeized gluon. It is the third bootstrap condition.

\section{Leading and next-to-leading orders}
It is well known that the bootstrap conditions described in the
preceding section are fulfilled in the leading order (LO) [5]. Indeed
the homogeneous LO Schroedinger equation has an eigenvalue
$E_0^{(1)}=-\omega^{(1)}(q)$ with the corresponding eigenfunction
$\Psi_0^{(0)}=c\chi$ (Eq. (6), $c=1/\sqrt{\langle\chi|\chi\rangle}$ is
the normalization factor):
\beq
H^{(1)}\chi=-\omega^{(1)}(q)\chi
\eeq
(upper indeces always mark the order in $\alpha_s$). One easily checks
that (24) is true with the explicit expressions (4)-- (8) for the
Hamiltonian and  $\chi$. Due to the trivial form of the latter
Eq. (24) is in fact a simple relation between the gluon trajectory and
the
gluonic interaction integrated over one of its momenta.

It can also be trivially shown that in the lowest order the impact
factors for any external particle reduce to $\chi$ as required.
Indeed (see[1])  in the LO for any particle
\beq
\Phi=-i\frac{\sqrt{N}}{2}\Gamma\chi.
\eeq
From this one gets
\beq
\langle\Psi_0^{(0)}|\Phi\rangle=
-i\frac{\sqrt{N}}{2}\Gamma\sqrt{\langle\chi|\chi\rangle}=
\frac{\sqrt{\pi\omega^{(1)}}}{q}\Gamma
\eeq
where we have used (5). Putting this in (23) we find that also the third
bootstrap condition
is satisfied in the LO, when we can neglect the power factor and 
take $\sin\pi\omega\simeq\pi\omega$

Now we pass to the next-to-leading order (NLO). Again we have to fulfil
three different requirements. First, the spectrum of the Hamiltonian
should
include the gluon trajectory also in the NLO, Eq. (19)
We present the corresponding eigenfunction as
\beq
\Psi_0=c\chi+\Psi_0^{(1)} 
\eeq
where $\chi$ is the known LO eigenfunction and the second term is the
(unknown) NLO contribution. Separating perturbation orders in the
Hamiltonian and in the eigenvalue $E_0=-\omega(q)$ and taking the NLO  we
get an equation for $\Psi_0^{(1)}$ 
\beq
(H^{(1)}+\omega^{(1)}(q))\Psi_0^{(1)}=
-c(H^{(2)}+\omega^{(2)}(q))\chi
\eeq
For this equation to be soluble, the inhomogeneous term should be
orthogonal to the solution of the corresponding homogeneous equation, that
is, to $\chi$, recall Eq. (24). So we get a condition
\beq
\langle\chi|H^{(2)}+\omega^{(2)}(q)|\chi\rangle=0
\eeq
With $\chi$ a known function, this equation is in fact a relation
between the gluonic NLO interaction in the colour channel $R=g$ and the
NLO Regge trajectory, integrated over the initial and final gluon
relative momenta. This relation was first obtained in [1] from different
arguments. Note that in contrast to the Eq. (19),
valid for any $q$ and $q_1$, relation (29) is only an identity in
$q$. So its contents is much weaker than that of (19). For this reason we 
shall call it a weak bootstrap contition, leaving the definition of the
strong bootstrap condition for the equations (24) in the LO and
(19) in the LO and NLO.

Now we come to the second condition, which is that also in the NLO there
should be no contribution to the absorptive part other than from the
gluonic Regge pole. However this condition is trivially staisfied in the
NLO. Indeed matrix elements
\[ \langle\Phi_{p.t}|\Psi_n\rangle, \ \ n\neq 0\]
with eigenfunctions different from $\Psi_0$ are all zero in the lowest
order. So they have at least order $\alpha_s$. The representation (13)
contains products of two such matrix elements. So the part of (13) which
comes from the states other than $\Psi_0$ is at least two orders smaller
than the leading term (of the order $\alpha_s^2$). Therefore it does not
contribute in the NLO at all. Thus in the NLO, irrespective of the form of
the
impact factors, there is no contribution to the absorptive part
of the amplitude from states other than a single reggeized gluon.

It is remarkable that this is only true for the absorptive part of the
scattering amplitude for real particles. If one considers the particle-
reggeon amplitudes instead, then  admixture of other
intermediate states, apart from a single reggeized gluon, seems
possible.
Indeed take $\Psi_t(E)$ as an example. One immediately sees from Eq. (18)
that terms with $n\neq 0$ will generally give a nonzero contribution in
the NLO unless the impact factor  coincides with the eigenfunction
$\Psi_0$ also in the the NLO (which seems improbable for every possible
external
particle).
This
contribution disappears when one takes the product in (17) due to 
orthogonality of the eigenfunction with $n=0$ to all others. However in 
the particle-reggeon amplitude $\Psi_t(E)$ itself the NLO contribution
from higher gluonic states may be present.

This circumstance makes one think about the presence of such states in the
reggeon-reggeon amplitudes which enter the unitarity relation for the
absorptive part of the initial amplitude. Should such states be present,
it would invalidate the derivation of the BFKL equation in the NLO. 

The form which takes the third bootstrap condition (23) in the NLO was
obtained in [1]. As far as we know it has not been checked for any
particle so far. 

\section{Solution of the strong bootstrap condition}
In this section we are going to demonstrate that the anzatz proposed in
[6] and used in [2] actually solves the strong bootstrap relation (19)
in the LO and NLO orders. The ansatz  introduces 
a single function $\eq$ through each both the potential and the gluon
Regge 
trajectory are expressed.
In [2] we used an unsymmetric potential $W$  defined as follows
\beq
W(q_1,q'_1)=\left(\frac{\ea}{\ec}+\frac{\eb}{\ed}\right)\frac{1}{\ee}-
\frac{\eq}{\ec\ed}
\eeq
The trajectory is  expressed via $\eq$ as
\beq
\omega(q)=-\int d^{D-2}q_1\frac{\eq}{\ea\eb}
\eeq
This anzatz guarantees fulfilment of the bootstrap relation in the
form
\beq
\int d^{D-2}q'_1W(q_1,q'_1)=\oq-\oa-\ob
\eeq
In the LO one has
\beq
\eta^{(0)}(q)=q^2/a
\eeq
(see Eq. (7)).

To pass to the symmetric  potential $V$ one has first separate
from $W$ the integration measure factor $1/(\ec\ed)$ and then symmetrize
substituting this factor for $1/\sqrt{\ea\eb\ec\ed}$. This gives the
desired relation
\beq
W(q_1,q'_1)=\sqrt{\frac{\ea\eb}{\ec\ed}}V(q_1,q'_1)
\eeq
The bootstrap relation (32) then transforms into
\beq
\int d^{D-2}q'_1V(q_1,q'_1)\frac{1}{\sqrt{\ec\ed}}=(\oq-\oa-\ob)
\frac{1}{\sqrt{\ea\eb}}
\eeq
This is nothing but the equation (19) for the gluon Regge pole
solved by
\beq
\Psi_0(q_1)=c\frac{1}{\sqrt{\ea\eb}}
\eeq
So our anzatz just solves the strong bootstrap relation.

To compare our results with [3] we have finally  to relate our
symmetric potential $V$ to the
irreducible kernel $K_r$ used in [1,3]. The latter is defined in
respect to the metric with a factor $q_1^2q_2^2$ in the denominator.
Therefore identification with $V$ requires passing to our metric.
We then find
\beq
V(q_1,q'_1)=\frac{K_r(q_1,q'_1)}{\sqrt{q_1^2q_2^2{q'_1}^2{q'_2}^2}}
\eeq

Combining (34) and (37) we find the final relation of the irreducuble
kernel
$K_r$ of [1,3] and the unsymmetric potential $W$ of [2]
\beq
K_r(q_1,q'_1)=\sqrt{q_1^2q_2^2{q'_1}^2{q'_2}^2}
\sqrt{\frac{\ec\ed}{\ea\eb}}W(q_1,q'_1)
\eeq
This  relation has not, in all probability, been taken into account in
[3] when
discussing our results. In the following section we shall show that
with the form of $\eq$ matched to the quark contribution to the gluon
Regge trajectory one obtains the correct form of the quark contribution
to the kernel $K_r$.

\section{The quark contribution to the trajectory and potential}
We are going to show that, first, the quark contribution to the gluon
Regge trajectory has the form implied by our anzatz (31). We
shall determine the part of $\eq$ coming from this contribution.
Then using (30) and (38)) we shall find the corresponding part of the
irreducible
kernel $K_r$ and compare it with the results of [3]. 

The part of the gluon trajectory which comes from quark is given by
the expression [3]
\beq
\omega_Q^{(2)}=2bq^2\int\frac{d^{D-2}q_1}{q_1^2q_2^2}(q^{2\epsilon}-
q_1^{2\epsilon}-q_2^{2\epsilon}),
\eeq
where
\beq
b=\frac{g^4NN_F{\rm \Gamma}(1-\epsilon){\rm \Gamma}^2(2+\epsilon)}
{(2\pi)^{D-1}(4\pi)^{2+\epsilon}\epsilon{\rm \Gamma}(4+2\epsilon)}.
\eeq
On the other hand, presenting
\beq
\eq=\eta^{(0)}(q)(1+\xq)
\eeq
we have in the NLO from (31)
\beq
\omega^{(2)}(q)=-aq^2\int\frac{d^{D-2}q_1}{q_1^2q_2^2}(\xq-
\xa-\xb)
\eeq
As we observe, the form of (39) follows this pattern. Comparing (39) and
(42)
we identify
\beq
\xi_Q(q)=-\frac{2b}{a}q^{2\epsilon}
\eeq

Now we pass to the irreducible kernel $K_r$. From (30) and (38) we
express it
via $\xq$ as
\[
K^{(2)}_r(q_1,q'_1)=\frac{1}{2}a\sqrt{q_1^2q_2^2{q'_1}^2{q'_2}^2}
\Big[\frac{1}{k^2}\sqrt{\frac{q_1^2{q'_2}^2}{{q'_1}^2q_2^2}}
(\xa+\xd-\xc-\xb-2\xe)
+\]\beq
\frac{1}{k^2}\sqrt{\frac{q_2^2{q'_1}^2}{{q'_2}^2q_1^2}}
(\xb+\xc-\xa-\xd-2\xe)
-\frac{q^2}{\sqrt{q_1^2q_2^2{q'_1}^2{q'_2}^2}}
(2\xq-\xa-\xb-\xc-\xd)\Big]
\eeq
Putting here the found $\xq$, Eq. (43), we obtain the quark contribution
\[
K^{Q}_r(q_1,q'_1)=-b
\Big[\frac{q_1^2{q'_2}^2}{k^2}
(q_1^{2\epsilon}+{q'_2}^{2\epsilon}-q_2^{2\epsilon}-{q'_1}^{2\epsilon}
-2k^{2\epsilon})+\]\beq
\frac{q_2^2{q'_1}^2}{k^2}
(q_2^{2\epsilon}+{q'_1}^{2\epsilon}-q_1^{2\epsilon}-{q'_2}^{2\epsilon}
-2k^{2\epsilon})
-q^2(2q^{2\epsilon}-q_1^{2\epsilon}-q_2^{2\epsilon}-{q'_1}^{2\epsilon}
-{q'_2}^{2\epsilon})
\Big]
\eeq
Comparing this expression with the one found in [3] (Eq. (47) of that
paper) we observe that they are identical. This means that our bootstrap
condition and the anzatz to solve it are valid at least for the quark
contribution in the NLO.

\section{Conclusions}
Comparing our bootstrap
results [2] for the gluonic interaction in the
gluon colour
channel with the direct calculations of its quark part in [3] we find a
complete agreement. This leaves a certain dose of optimism as to
the total potential calculated in [2] being the correct one.

\section{Acknowledgements}
The author expresses his deep gratitude to Profs. Bo Anderson and
G.Gustafson for their hospitality during his stay at Lund University,
where these comments were written.

\section{References}

1. V.S.Fadin and R.Fiore, hep-ph/9807472.

\noi 2. M.A.Braun and G.P.Vacca, hep-ph/9810454.

\noi 3. V.S.Fadin, R.Fiore and A.Papa, hep-ph/9812456.

\noi 4. V.S.Fadin, E.A.Kuraev and L.N.Lipatov, Phys. Lett. {\bf B60}
(1975)
50.

\noi 5. L.N.Lipatov, Yad. Fiz. {\bf 23} (1976) 642.

\noi 6. M.A.Braun, Phys. Lett. {\bf B345} (1995) 155; {\bf B348} (1995)
190.
 \end{document}